\documentclass[aps,prl,showpacs,superscriptaddress,twocolumn,final,numerical,10pt]{revtex4-1}
\usepackage{amsmath,amsfonts,amsthm}

\usepackage{graphicx,color}
\usepackage{multirow}
\usepackage{braket}
\usepackage{bbm}
\usepackage{textcomp}

\newcommand{\ie}{\textit{i.e.}}
\newcommand{\eg}{\textit{e.g.}}

\DeclareMathOperator{\tr}{tr}
\DeclareMathOperator{\Prob}{Prob}

\begin{document}
\title{Certifying experimental errors in quantum experiments}

\author{Tobias~Moroder}
\affiliation{Naturwissenschaftlich-Technische Fakult\"at, Universit\"at Siegen, Walter-Flex-Str.~3, D-57068 Siegen, Germany}
\affiliation{Institut f\"ur Quantenoptik und Quanteninformation, \"Osterreichische Akademie der Wissenschaften, Technikerstr.~21A, A-6020 Innsbruck, Austria}
\author{Matthias~Kleinmann}
\affiliation{Naturwissenschaftlich-Technische Fakult\"at, Universit\"at Siegen, Walter-Flex-Str.~3, D-57068 Siegen, Germany}
\author{Philipp~Schindler}
\author{Thomas~Monz}
\affiliation{Institut f\"ur Experimentalphysik, Universit\"at Innsbruck, Technikerstr. 25, A-6020 Innsbruck, Austria}
\author{Otfried~G\"uhne}
\affiliation{Naturwissenschaftlich-Technische Fakult\"at, Universit\"at Siegen, Walter-Flex-Str.~3, D-57068 Siegen, Germany}
\affiliation{Institut f\"ur Quantenoptik und Quanteninformation, \"Osterreichische Akademie der Wissenschaften, Technikerstr.~21A, A-6020 Innsbruck, Austria}
\author{Rainer~Blatt}
\affiliation{Institut f\"ur Quantenoptik und Quanteninformation, \"Osterreichische Akademie der Wissenschaften, Technikerstr.~21A, A-6020 Innsbruck, Austria}
\affiliation{Institut f\"ur Experimentalphysik, Universit\"at Innsbruck, Technikerstr. 25, A-6020 Innsbruck, Austria}

\begin{abstract}
When experimental errors are ignored in an experiment, the subsequent analysis of its results becomes questionable. We develop tests to detect systematic errors in quantum experiments where only a finite amount of data is recorded and apply these tests to tomographic data taken in an ion trap experiment. We put particular emphasis on quantum state tomography and present three detection methods: the first two employ linear inequalities while the third is based on the generalized likelihood ratio.
\end{abstract}

\pacs{03.65.Ta, 03.65.Wj, 06.20.Dk, 42.50.Dv}

\maketitle

\emph{Introduction.}---Measurements are central to acquiring information about the underlying system in any quantum experiment. However, for quantum systems of increased complexity, the analysis of all measurement data gets challenging when one deals with both statistical and systematic errors. Statistical errors refer to the intrinsic problem that true probabilities are never accessible in any experiment but are merely approximated from count rates which lead to relative frequencies. A well-known example where statistical effects play a dominant role is quantum state tomography~\cite{qse_book}: the task to determine an unknown state by means of appropriate measurements. Here the deviations between probabilities and relative frequencies cause severe problems in the actual state reconstruction, since na\"ively using the frequencies in Born's rule easily leads to unphysical ``density operators,'' meaning that some eigenvalues are negative. This problem can be circumvented by reconstruction principles that explicitly account for statistical effects~\cite{hradil97a,james01a}.

The analysis is generally further complicated because of additional systematic errors, \eg, caused by drifts in the state generation, misalignment in the measurements or fluctuations of external parameters. To reconstruct the state from the observed data one requires an operator assignment for each classical outcome of the performed measurements. This measurement model is essential, not just for quantum state tomography, but also to certify state characteristics like entanglement via entanglement witnesses~\cite{lewenstein00a,terhal00a} or  applications as quantum key distribution to prove security in the calibrated device scenario~\cite{scarani09a}. However, in a real experiment the measured observables might deviate from this employed description due to systematic errors. This mismatch can have severe impact on the analysis and can lead to, for instance, spurious entanglement detection as exemplified in Ref.~\cite{acin06a} or insecurity in quantum key distribution~\cite{qi07a,lydersen10a}. Though deviations of this kind have been discussed and partially countermeasured by different techniques~\cite{lougovski09a,moroder10a,bancal11a,moroder12a,rosset12a}, it has not yet been investigated how to distinguish them from statistical errors. An exception is Ref.~\cite{schwarz11a}, where drifts in the source are detected by measurements on subsequent states.

In this Letter we present experimentally and theoretically three methods to detect whether systematic errors are statistically significant, \ie, if there is merely a small probability that the observed results were generated by statistical effects only. In that case, the model becomes questionable and further analysis must involve a refined model or include other means of treating systematic errors. We emphasize that the techniques outlined below can only falsify but never verify that systematic errors are absent. Some errors, as, for example, depolarizing noise, are not detectable without further calibration. Still, we recommend that these tests are applied before reconstructing actual quantum states since they serve as additional systematic error checks after calibrating the setup. Three procedures are presented in detail, the first two use linear inequalities that are satisfied if no systematic errors are present, while the third is based on the likelihood ratio~\cite{knight,blume10b}. Note that other techniques from hypothesis testing, like the prediction-based-ratio analysis~\cite{zhang11a} or the chi-square goodness-of-fit~\cite{mood}, provide alternative procedures to test for systematic errors.

\textit{Tomography setting.}---A common tomography protocol uses $3^n$ possible combinations of Pauli operators on $n$ qubits and one measures locally the respective expectation values in the associated eigenbasis which provides $2^n$ distinct outcomes, yielding a total of  $3^n \times 2^n = 6^n$ different outcomes. Note that an $n$-qubit density operator is already determined by $4^n - 1$ parameters, \ie, this measurement scheme collects an overcomplete data set. This tomography protocol is known as the Pauli measurement scheme~\cite{james01a} which has been used for $n$-qubit systems in ion traps~\cite{haeffner05a} or photonic setups~\cite{kiesel07a}.

More generally, we consider a tomography protocol with measurements for different settings labeled by $s$ and which registers the respective frequencies $f_k^s = m^s_k / N_{\rm s}$, where $m^s_k$ denotes the counts of the specific outcome $k$ in $N_{\rm s}$ repetitions of this experiment. The repetitions $N_{\rm s}$ are assumed to be equal for each setting. The observables $M_k^s$ are the attributed measurement operators and they span the complete operator space in order to enable a full reconstruction of the density operator. Most often this set is overcomplete, \ie, the operators are not independent of each other which can be expressed in terms of linear identities $\sum c_k^s M_k^s = 0$ using real coefficients $c_k^s$. The set of probabilities consistent with this quantum model are all distributions $P_{\rm qm}(k|s)=\tr(\rho M_k^s)$ that can be written using a density operator $\rho$.

\textit{Witness test.}---The set of distributions consistent with the assumed quantum model can be characterized by linear inequalities. This is in analogy to entanglement witnesses~\cite{horodecki96b,terhal00a} for separable states or Bell inequalities~\cite{peres99a} for local hidden variable models. Consider a set of real coefficients $w=w_k^s$ that define a positive semidefinite operator via $\sum w_k^s M_k^s=Z_w \succeq 0$, \ie, all eigenvalues are non-negative. Then for each such $w$ the expectation value of any probability distribution from the quantum model $P_{\rm qm}$ satisfies
\begin{equation}
\label{eq:witness}
w \cdot  P_{\rm qm} \equiv \sum_{s,k} w_k^s P_{\rm qm}(k|s) = \tr (\rho Z_w)\geq 0.
\end{equation}
Thus a distribution $P$ with $w \cdot P < 0$ is incompatible with the assumed quantum model, and any such distribution can be detected by a set of coefficients $w$ of the described form (even with partial information~\cite{moroder08a}). Thus we refer to $w$ as a witness for systematic errors, but note that its associated operator $Z_w$ is not an entanglement witness.

Equation~(\ref{eq:witness}) is formulated on the level of probabilities which are not accessible in the experiment. Nevertheless one can replace the probabilities by the observed frequencies $f=f_k^s$ and consider the sample mean $w \cdot f \equiv \sum w_k^s f_k^s$ of the witness. Then $w \cdot f \geq 0$ does not need to hold anymore because statistical effects can produce a negative value. However, the probability to observe large deviations from the true mean is bounded and decreases exponentially with the number of performed repetitions. A quantitative statement is given by Hoeffding's tail inequality \cite{hoeffding}, as similarly used for example in efficient fidelity estimation~\cite{flammia11a,silva11a}. We emphasize that this inequality is even valid for small data sets containing only few or no counts for certain outcomes.

\textit{Proposition~$1$.}---Consider a witness $w=w_k^s$ obeying $\sum w_k^s M_k^s=Z_w \succeq 0$.  If the data are generated by the quantum model $P_{\rm qm}(k|s)=\tr(\rho M_k^s)$, then for all $t > 0$,
\begin{equation}
\label{eq:statement1}
\Prob[ w \cdot f \leq -t] \leq \exp ( -2 t^2 N_{\rm s}/C_w^2 )
\end{equation}
with $C_w^2 = \sum_s \left( w_{\rm max}^s - w_{\rm min}^s \right)^2$, where $w_{\rm max/min}^s$ are the optima for setting $s$ over all outcomes $k$. A proof is given in the appendix.

The interpretation is as follows: Suppose that one carries out an experiment for a previously chosen witness $w$ and fixed error probability $\alpha$, which one still tolerates before one announces a systematic error. Using Proposition~$1$ one computes the necessary violation $t_{\alpha}=\sqrt{-C^2_w \log(\alpha) / 2 N_{\rm s}}$. If one now registers frequencies $f_{\rm obs}$ with $w \cdot f_{\rm obs} \leq -t_{\alpha}$, then the probability that any error-free experiment would produce such data is less than $\alpha$ and one says that a systematic error is significant at significance level $\alpha$. However, for given data it is more common to report the smallest $\alpha$ such that the systematic error is significant. This is also called the p-value in hypothesis testing~\cite{knight}. Proposition~$1$ states that this p-value has an upper bound of $\exp[-2 (w\cdot f_{\rm obs})^2 N_{\rm s}/C_w^2 ]$ if $w\cdot f_{\rm obs} < 0$.

\begin{figure}[t]
\centering
\includegraphics[width=0.76\columnwidth]{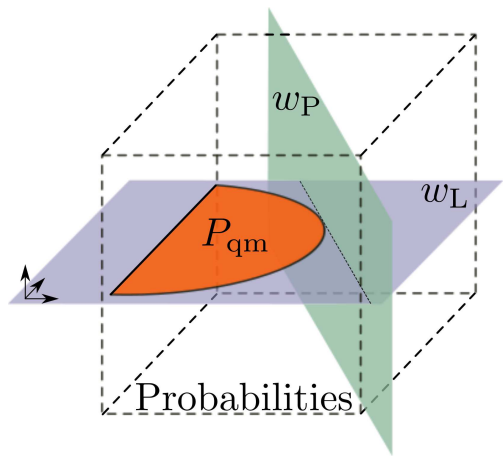}
\caption{The admissible probabilities from the quantum model $P_{\rm qm}$ typically form a convex, lower dimensional subset within all possible probability distributions (dashed cube). This dimension reduction stems from additional linear relations that a probability distribution from the quantum model must fulfil. These relations are checked by witnesses $w_{\rm L}$, while $w_{\rm P}$ verify positivity of the density operator.}
\label{fig:witness}
\end{figure}

\textit{Witness structure.}---Each witness $w$ as defined above can be decomposed into two conceptually different parts. One that solely verifies positivity of an underlying density operator, denoted as $w_{\rm P}$, and into another part $w_{\rm L}$ that only checks the linear dependencies within the assumed measurement operators, such that one obtains $w=w_{\rm P}+ w_{\rm L}$. It turns out that these two parts of the witness are orthogonal. Note that the witness $w_{\rm P}$ uniquely describes the operator $\sum w_{{\rm P} k}^{\phantom{{\rm P}}s} M_k^s=Z_w$, while the witness $w_{\rm L}$ (and also $-w_{\rm L}$) vanishes due to the linear relations $\sum w_{{\rm L}k}^{\phantom{{\rm L}}s} M_k^s = 0$. Figure~\ref{fig:witness} gives a schematic picture of this situation.

\textit{Issue of negative eigenvalues.}---The above framework provides an answer to the issue of negative eigenvalues in linear inversion, since it is connected to witnesses of the type $w_{\rm P}$. Linear inversion refers to the state reconstruction process in which one estimates the unknown density operator by using the observed frequencies in Born's rule $\tr(\rho M_k^s)=f_k^s$. Since this set of linear equations is typically not exactly solvable because of overcompleteness one selects the operator $\rho_{\rm ls}$ which minimizes the least squares, $\sum [f_k^s - \tr(\rho_{\rm ls} M_k^s)]^2$. As one ignores the positivity constraint this operator $\rho_{\rm ls}$ will often represent an invalid density operator because some eigenvalues are negative, \ie, $\braket{\psi| \rho_{\rm ls} | \psi}< 0$.

\textit{Proposition~2.}---Let $\rho_{\rm ls}$ be the linear inversion using least squares and consider a given vector $\ket{\psi}$. If the data are generated by the quantum model $P_{\rm qm}(k|s)=\tr(\rho M_k^s)$, then for all $t >0$,
\begin{equation}
\Prob [ \braket{\psi| \rho_{\rm ls} | \psi} \leq - t \, ] \leq \exp ( -2 t^2 N_{\rm s} /C_w^2 )
\end{equation}
with $C_w^2$ as given in Proposition~$1$ computed from the unique $w_{\rm P}$ satisfying $\sum w_{{\rm P} k}^{\phantom{{\rm P}}s}M_k^s=\ket{\psi}\bra{\psi}$. A proof is given in the appendix.

This proposition shows that the probability to successfully guess a state $\ket{\psi}$, independently of the recorded data, where $\rho_{\rm ls}$ has a negative expectation value is exponentially suppressed.

\emph{Likelihood ratio test.}---In addition to the attributed quantum model $P_\mathrm{qm}(k|s)= \tr(\rho M_k^s)$ we can also describe the observations with a more general model assumption of independent distributions $P_\mathrm{ind}(k|s)=p_k^s \geq 0$ and $\sum_k p_k^s=1$ for each setting $s$. The question whether the observed data set is compatible with the assumed quantum model can now be addressed by comparing the maximal likelihoods of either model~\cite{knight}.

For that, we start from the likelihood for a distribution $P$ given the observed frequencies $f$, which is $ L(P)= \prod_{k,s} P(k|s)^{N_{\rm s} f^s_k}$
ignoring the multinomial prefactor. A quantum state $\rho_\mathrm{ml}$ that maximizes the likelihood $L(P)$ is considered to be a good estimate for the physical state \cite{qse_book,hradil97a}. In contrast, for the model with all independent distributions, the optimum is given by $p_k^s=f_k^s$. Since the quantum model is contained in this more general model, the likelihood of any quantum model can at best be equal to this optimal likelihood. Thus one finds $ L(f)\ge L[\tr(\rho_\mathrm{ml} M_k^s)]$ or equivalently, a non-negative log-likelihood ratio $\lambda_\mathrm{qm}= 2\log L(f)- 2\log  L[\tr(\rho_\mathrm{ml}M_k^s)]$.

The likelihood ratio test is based on the observation, that if the data are indeed generated from the quantum model then the probability for outcomes which satisfy $\lambda_{\rm qm}\ge t$ decreases rapidly if $t$ exceeds a certain value. Wilks' theorem~\cite{Wilks:1962} states that this ratio is distributed according to a chi-square distribution already for moderately large samples. However this theorem does not directly apply to $\lambda_{\rm qm}$ because of the positivity constraint; but it works for the slightly larger model where one performs the optimization (rather than over quantum models) over probabilities $P_{\rm nqm}(k|s)=\tr(X M_k^s)$ that can be written in terms of a Hermitian operator $X$. Note that $X$ can have negative eigenvalues, indicated by the subscript ``n'', while still obeying the positivity constraints $\tr(X M_k^s) \geq 0$ for the measurements $M_k^s$. With $X_\mathrm{ml}$ being a corresponding optimum we now study the log-likelihood ratio
\begin{equation}
 \label{eq:lambda_nqm}
 \lambda_{\rm nqm}= 2 \log L(f) - 2 \log L[\tr(X_\mathrm{ml}M_k^s)].
\end{equation}

\textit{Proposition~$3$.}---If the data are generated by the $d$-dimensional quantum model $P_\mathrm{qm}(k|s)= \tr(\rho M^s_k)$ with $K$ outcomes for each of the $S$ settings, then for all $ t>0$, as $N_{\rm s}\rightarrow \infty$,
\begin{equation}\label{me:wilks}
 \Prob[ \lambda_\mathrm{nqm}\geq t ]\rightarrow Q(\Delta/2,t/2),
\end{equation}
with the dimension deficit $\Delta=(K-1)S-(d^2-1)$ and the regularized incomplete gamma function $Q$~\footnote{Explicitly, $Q(s,x)= \int_x^\infty y^{s-1} \mathrm e^{-y} \mathrm dy/\int_0^\infty y^{s-1} \mathrm e^{-y} \mathrm dy$; the function $1-Q(\Delta/2,t/2)$ is the cumulative distribution function of a chi-square distribution with $\Delta$ degrees of freedom.}. A proof is given in the appendix.

The interpretation and application is analogous to Proposition~$1$. Though Proposition~$3$ is only a strict statement in the asymptotic case $N_{\rm s} \to \infty$, Eq.~\eqref{me:wilks} gives reliable values already for moderately large $N_{\rm s}$, as we will demonstrate below.

\emph{Experimental setup}---Experimentally, we study tomographic data from an ion trap quantum processor encoding qubits in the ground and the metastable state of $^{40}$Ca$^{+}$ ions where each ion represents a qubit. Details on the experimental setup can be found in Ref.~\cite{expsetup}. Single ions can be addressed with a tightly focused, off-resonant beam. Here the ac-Stark effect induces an operation of the form $\exp(-i \Omega_l \tau \sigma_{z,l} /2)$ on ion $l$, with the Rabi frequency $\Omega_l$ determined by detuning and intensity, and pulse duration $\tau$. Combined with collective, resonant operations on all qubits, state tomography according to the Pauli measurement scheme can be implemented on the trapped-ion quantum register.

In an experimental realization, the finite width of the focused beam results in residual ion-light interaction on next-neighbor qubits. The Rabi frequency of ion $k$ when addressing ion $j$ can be described by the addressing matrix $\Omega_{j,k}$. Thus the operation on the qubit register can then be written as $\exp(-i\sum_k\Omega_{j,k} \tau\sigma_{z,k}/2 )$. The addressing quality can be quantified with a cross-talk parameter $\epsilon = \max_{j \neq k}(\Omega_{j,k}/\Omega_{j,j})$, which can be increased by defocusing the addressed laser beam.

Using this setup we perform tomography on various states and investigate whether the obtained data suffer from any kind of systematic errors. This includes data for Greenberger-Horne-Zeilinger states on $4$ ions, $\ket{GHZ}=(\ket{0000}+\ket{1111})/\sqrt{2}$, where we intentionally increased the cross-talk $\epsilon$ to test the presented techniques, a large data set on a two-qubit Bell state $\ket{\psi^-}=(\ket{01}-\ket{10})/\sqrt{2}$ and measurements on the ground state $\ket{SSSS}=\ket{1111}$. Moreover we re-analyse observations on a W-state on $5$ qubits, $\ket{W}=(\ket{00001}+\ket{00010}+\ket{00100}+\ket{01000}+\ket{10000})/\sqrt{5}$ and a bound-entangled (BE) Smolin state~\cite{bound_entanglement_uibk}.

\emph{Empirical findings.}---
\begin{table}[t]
\begin{tabular}{l|r|r|r || c | c || c | c}
State&$n$&
\multicolumn{1}{|c|}{$N_{\rm s}$}& $\epsilon\phantom{aa}$ &$w_{\rm L}$&$w_{\rm P}$&LR&LR$^*$\\\hline\hline
\multirow{3}{*}{GHZ}&\multirow{3}{*}{4}
      &  750&$0.20\;^\star$& $97\%$ & $10^{-6} \%$ & $10^{-10}\%$ & $10^{-9} \%$  \\
     &&  750&$0.12\;^\star$& $100\%$          & $ 10^{-7}\%$ & $0.024 \%$&$ 0.14 \%$      \\
     &&  600& $<0.03$ & $79 \%$ & $81 \%$ & $0.91 \%$ & $ 4.1 \%$     \\
   \hline
Bell&2&61650& $<0.03$ & $100\%$          &  $100\%$        & $50 \%$   & $49 \%$      \\
   \hline
SSSS&4& 2600& $<0.03^\dag$    & $48 \%$  &  $84 \%$ & $0.037 \%$   & $0.008 \%$      \\
   \hline
BE  &4& 5200& $<0.03$ & $99 \%$ & $14 \%$ & $35 \%$   & $36 \%$      \\
   \hline
W   &5&  100& $0.04$      & $49 \%$ & $91 \%$ & ($0.081 \%$) & $5.5 \%$     \\
\end{tabular}
\caption{\label{mt:pvals}
Systematic error analysis for various experimental data according to different tests, \ie, $w_{\rm L}$ and $w_{\rm P}$ refer to the witness test, while LR corresponds to the likelihood ratio test. The values are upper bounds on the p-value of each test. The specifications are: $n$ number of qubits, $N_{\rm s}$ number of measurements per setting, $\epsilon$ measured cross-talk parameter.
LR$^*$ is obtained using a parametric bootstrapping method~\cite{bootstrapping} with $1000$ samples. For data marked with $^\star$ we manually increased the cross-talk, while $^\dag$ have been intensity fluctuations.}
\end{table}%
At first we implement the witness test, see Table~\ref{mt:pvals}. Let us stress that Proposition~$1$ does not allow us to determine and to evaluate the witness $w$ from the same data. If one would do so then one effectively employs $\min_{w} w \cdot f$ instead of $w \cdot f$ as required in Proposition~$1$. Because of that we divide the observed data into two equally sized parts, yielding frequencies $f_1$ and $f_2$. Afterwards we use the first part $f_1$ to determine a reasonable witness $w$, which is evaluated on the second part, $w \cdot f_2$. Here we choose either of the two types of witnesses testing positivity $w_{\rm P}$ or linear dependencies $w_{\rm L}$. As witness $w_{\rm P}$ we select the witness that corresponds to the projector onto the smallest eigenvalue on the linear inversion $\rho_\mathrm{ls}$ using the first data set $f_1$. For the linear dependencies we use $w_{\rm L} = -f_1 +\tr(\rho_{\rm ls} M_k^s)$, because it gives the largest negative expectation value $w_{\rm L} \cdot f_1$ on the first data. Note that the employed choices are not necessarily optimal~\cite{jungnitsch10a}. If the observed value $w\cdot f_2$ is negative, we ask for the statistical significance as explained after Proposition~$1$. If we choose a significance level of for instance $\alpha = 0.1\%$, the witness $w_{\rm P}$ reliably detects the artificially introduced cross-talk for the GHZ-state experiments, while $w_{\rm L}$ is less powerful for these examples.

\begin{figure}[t]
\includegraphics[width=0.87\columnwidth]{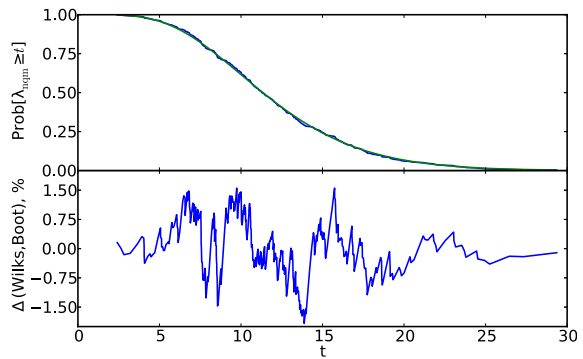}
\caption{\label{mf:lrd}
Fraction of runs with a log-likelihood ratio $\lambda_{\rm nqm} \geq t$ from a 411-fold repetition of the Bell-state experiment.  In the upper graph, the shaky blue line corresponds to the experimental data, while the smooth green line is the prediction according to Wilks' theorem. The lower graph shows the difference between both curves.}
\end{figure}

The likelihood ratio test, as a third method, is best suited for a larger number of samples, since Proposition~$3$ makes only a strict statement for $N_{\rm s}\rightarrow \infty$.
In Figure~\ref{mf:lrd} we compare the empirical distribution between a two-qubit Bell experiment using $150$ samples per setting and the predicted distribution according to Wilks' theorem. Hence for the two-qubit case this number might already be sufficiently close to this limit. This observation is further supported by a comparison with a bootstrapping method~\cite{bootstrapping} (see appendix) which produces similar results as the ones obtained from Proposition~$3$. Based on these observations we are confident that the results using Proposition~$3$ for finite $N_{\rm s}$ are trustworthy for all data from Table~\ref{mt:pvals} except for the W-state, which has a too low number of samples. Evaluating the experimental data we detect again the manually increased cross-talk in the GHZ experiments, but now also some discrepancies in the SSSS experiment, which occurred because of intensity fluctuations during the experiment~\footnote{The optimizations are done using the solver \texttt{cp} from the Python package CVXOPT.}.

\textit{Conclusion and outlook.}---Tomographic reconstruction of quantum states can be problematic since nonphysical properties, such as negative eigenvalues, might occur. One possible solution is to use reconstruction schemes, which by construction result in a valid state. Then, however, serious concerns remain, since negative eigenvalues can also be a signature of systematic errors. We have provided tests which can be used to distinguish systematic from statistical errors in quantum experiments. These tests were shown to recognize systematic errors in real tomographic data from ion trap experiments.

Though we formulated our result for the case of state tomography, our methods can be applied to other assumptions like the nonsignaling condition in Bell experiments. From the more general perspective, many experiments in physics aim to determine parameters in an assumed theoretical model. Our results show that it is possible to give rigorous estimates on whether the assumed model class is inappropriate.

\begin{acknowledgments}
We thank K.M.R. Audenaert, K. Banaszek, M. Cramer, O. Gittsovich, S. Glancy, M. Guta, D. Gross, B. Jungnitsch, H. Kampermann, E. Knill, S. Niekamp and Y. Zhang for stimulating discussions. The authors are grateful for the opportunity to participate in the Heraeus Summer School ``Modern Statistical Methods in QIP'', where parts of this research have been developed. This work has been supported by the Austrian Science Fund FWF (START prize No. Y376-N16, SFB FoQuS No. F4002-N16), the EU (AQUTE, Marie Curie CIG 293993/ENFOQI), the BMBF (CHIST-ERA network QUASAR), the Institut f\"ur Quanteninformation GmbH and by the European Research Council. This research was funded by the Office of the Director of National Intelligence (ODNI), Intelligence Advanced Research Projects Activity (IARPA), through the Army Research Office grant~W911NF-10-1-0284. All statements of fact, opinion or conclusions contained herein are those of the authors and should not be construed as representing the official views or policies of IARPA, the ODNI, or the U.S. Government.
\end{acknowledgments}

\appendix

\section{Appendix}

\textit{Proof of Proposition~$1$.}---The proposition uses Hoeffding's tail inequality~\cite{hoeffding} and the property that valid quantum distributions $P_{\rm qm}$ have a non-negative expectation value $w \cdot f \geq 0$ due to Eq.~\eqref{eq:witness}. Hoeffding's inequality states that the sample mean $\bar X= \sum X_i /N$ of $N$ independent, not necessarily identical distributed, bounded random variables $X_i$ with $\Prob[X_i \in [a_i, b_i]]=1$ for $i=1,\dots N$ satisfies
\begin{equation}
\Prob[ \bar X - \mathbbm{E}(\bar X) \leq -t] \leq \exp[-2N^2t^2/\sum (b_i-a_i)^2]
\end{equation}
for all $t > 0$ and $\mathbbm{E}(\bar X)$ denoting the mean value of $\bar X$. In order to prove the proposition we identify $\bar X$ with the sample mean of the witness. This is achieved as follows:

Suppose that $Y_i^s$ denotes the random variable associated with the $i$-th repetition of the measurement setting $s$. In case of the measurement outcome $k$, $Y^s_i$ takes the value $S w_k^s$ where $S$ denotes the total number of measurement settings. It is bounded between $Y_i^s \in  [S w^s_{\rm min}, S w^s_{\rm max}]$. Then the sample mean of all these variables
\begin{equation}
\bar Y =\frac{1}{S N_{\rm s}} \sum_{s,i} Y_i^s
\end{equation}
yields values
\begin{equation}
\frac{1}{S N_{\rm s}} \sum_{s,k} S w_k^s m_k^s= \sum_{s,k} w_k^s f_k^s=w \cdot f,
\end{equation}
where $m^s_k$ denotes the counts of the specific outcome $k$ in $N_{\rm s}$ repetitions of the measurement settings $s$.
 
Using Hoeffding's inequality together with the property that $\mathbbm{E}( \bar Y ) = w \cdot P_{\rm qm} \geq 0$ holds for any valid quantum distribution $P_{\rm qm}$ due to Eq.~\eqref{eq:witness} one arrives at
\begin{eqnarray}
\Prob[ w \cdot f \leq -t] &=& \Prob[ \bar Y \leq -t]  \\
&\leq& \Prob[ \bar Y - \mathbbm{E}(\bar Y) \leq -t] \\
&\leq & \exp[-2 t^2 N_{\rm s} /C_s^2],
\end{eqnarray}
with $C_s^2=\sum_s (w^s_{\rm max}-w^s_{\rm min})^2$. The first inequality holds because the set of all outcomes satisfying $\bar Y < -t$ is a subset of $\bar Y - \mathbbm{E}( \bar Y ) \leq -t$. This concludes the proof of the proposition.

\textit{Proof of Proposition~$2$.}---In order to prove the proposition we show $\braket{ \psi|\rho_{\rm ls}|\psi}=w_{\rm P} \cdot f$, which can then be used in Proposition~$1$ to obtain the final statement.

Given a valid decomposition $w$ satisfying $\sum w_k^{s} M_k^s=\ket{\psi}\bra{\psi}$ one obtains
\begin{eqnarray}
\braket{\psi| \rho_{\rm ls}|\psi} &=& \sum w_k^s \tr(M_k^s \rho_{\rm ls})  = \sum w_k^s f_{{\rm P}k}^{\phantom{{\rm P}}s}\\
&=& w \cdot f_{\rm P} = w_{\rm P} \cdot f,
\end{eqnarray}
together with the solution of least squares $\tr(M_k^s \rho_{\rm ls})=f_{{\rm P}k}^{\phantom{{\rm P}}s}$.

\textit{Proof of Proposition~$3$.}---We start by a rough statement of Wilks' theorem (cf.\ \eg, {13.8.1} in Ref.~\cite{Wilks:1962}). Suppose that we have a family of models, that is sufficiently smoothly parameterized by an open set $\Omega\subset \mathbb R^r$ and let $A$ be a subspace of $\mathbb R^r$ of dimension $r-\Delta$. If we draw  $N'$ samples from our model with some (unknown) parameter $z\in A\cap \Omega$, then the distribution of the log-likelihood ratio
\begin{equation}
 \lambda_A = 2\log \sup_{x\in \Omega} L(x)- 2\log \sup_{y\in A\cap\Omega} L(y)
\end{equation}
converges to the $\chi^2_\Delta$ distribution as $N'\rightarrow \infty$. Hence, 
\begin{equation}
 \lim_{N'\rightarrow \infty} \Prob[\lambda_A \ge t] = Q(\Delta/2, t/2).
\end{equation}

For our model, the set of parameters $\Omega$ is given by $\tilde p^s_k$ with $k=1,\dotsc, K-1$, $s=1,\dotsc,S$ obeying $\tilde p^s_k>0$ and $\sum_k \tilde p^s_k <1$. Furthermore, we let $A$ be the subspace where $q^s_k = \tr(XM^s_k)$ for some Hermitian matrix $X$. We now basically arrived at Proposition~$3$. It only remains to change the notation from $\tilde p^s_k$ to $p^s_k$ via $p^s_K = 1- \sum_k \tilde p^s_k$ and to admit a more relaxed notation by essentially writing $\max_{x\in \overline \Omega}$ instead of $\sup_{x\in \Omega}$.

\textit{Bootstrapping for Table~\emph{I}.}---We first calculate the maximum likelihood state $\rho_\mathrm{ml}$ from the measured data and then simulate the tomographic process using that state. From the simulated data we then calculate the log-likelihood ratio $\lambda_\mathrm{nqm}$. We repeat this procedure 1000 times and compare the distribution with the predicted chi-square distribution. In our examples, the distribution matches always almost perfectly, however with a slightly different parameter $\Delta$. We then determine $\Delta'$ to be such, that $Q(\Delta'/2, m) = 1/2$, where $m$ is the median of the sampled distribution. Using $\Delta'$, we obtain the values of the column LR$^*$ of Table~\ref{mt:pvals}.

\clearpage

\end{document}